\documentclass[pra,twocolumn,floatfix,superscriptaddress,longbibliography,notitlepage]{revtex4-2}
\usepackage{amssymb,amsmath,amsthm,color,graphicx,times,graphicx}
\usepackage{caption}
\usepackage{ragged2e}
\DeclareCaptionJustification{justified}{\justifying}
\usepackage{hyperref}
\usepackage{braket}
\usepackage{graphicx}
\usepackage{dcolumn}
\usepackage{appendix}
\usepackage{subcaption}
\usepackage{bm}

\providecommand{\openone}{\leavevmode\hbox{\small1\kern-4.3pt\normalsize1}}

 \usepackage{orcidlink}

\theoremstyle{plain}

\theoremstyle{definition}

\begin{document}
\title{Hermitian vs non-Hermitian quantum thermometry}

\author{Anass Hminat \orcidlink{0009-0007-3677-3952}}\affiliation{LPHE-Modeling and Simulation, Faculty of Sciences, Mohammed V University in Rabat, Rabat, Morocco.}
\author{Abdallah Slaoui \orcidlink{0000-0002-5284-3240}}\affiliation{LPHE-Modeling and Simulation, Faculty of Sciences, Mohammed V University in Rabat, Rabat, Morocco.}\affiliation{Centre of Physics and Mathematics, CPM, Faculty of Sciences, Mohammed V University in Rabat, Rabat, Morocco.}
\author{Rachid Ahl Laamara \orcidlink{0000-0002-8254-9085}}\affiliation{LPHE-Modeling and Simulation, Faculty of Sciences, Mohammed V University in Rabat, Rabat, Morocco.}\affiliation{Centre of Physics and Mathematics, CPM, Faculty of Sciences, Mohammed V University in Rabat, Rabat, Morocco.}\author{Mourad Telmini \orcidlink{0000-0001-7428-6468}}\affiliation{LSAMA, Department of Physics, Faculty of Science of Tunis, University of Tunis El Manar, 2092 Tunis, Tunisia}

\begin{abstract}
We investigate the dephasing dynamics of a qubit as an effective mechanism for estimating the temperature of its surrounding environment for different symmetrizes. Our approach is fundamentally quantum, leveraging the qubit's susceptibility to decoherence without necessitating thermal equilibrium with the system under study. We also examine how symmetry properties affect the accuracy of information retrieval and the robustness of quantum information storage in such systems, highlighting their potential advantages in mitigating decoherence effects. The optimization of quantum Fisher information is performed with respect to both the interaction duration and the environmental temperature, focusing on Ohmic-like spectral density environments. Furthermore, we explicitly identify the optimal qubit measurement that attains the quantum Cramer-Rao bound for precision. Our findings reveal that optimal estimation arises from a complex interplay between the qubit's dephasing dynamics and the Ohmic characteristics of the environment with a particular focus on non-Hermitian systems that exhibit enhanced resilience to decoherence. Notably, optimal estimation does not occur when the qubit reaches a stationary state nor under conditions of complete dephasing.

\par
\vspace{0.25cm}
\end{abstract}
\date{\today}

\maketitle

\section{Introduction}
Quantum metrology, often referred to as quantum estimation theory, plays a critical role in advancing high-precision measurement technologies across diverse scientific fields \cite{Braunstein1996}. Its primary aim is to enable measurements with exceptional accuracy. In the classical domain, foundational contributions date back to the 1940s, with the independent works of Cramér \cite{Cramer1946} and Rao \cite{Rao1945}, who established a fundamental limit on the variance of any estimator. This limit, later generalized to multiparameter scenarios by Darmois \cite{Darmois1945}, is known as the Cramér-Rao bound and is intrinsically linked to the concept of Fisher information, introduced by Fisher in the 1920s \cite{Fisher1923}. Fisher information is a cornerstone of estimation theory, and its optimization over all possible quantum measurements yields the quantum Fisher information \cite{Braunstein1994}, which sets a quantum analogue to the Cramér-Rao bound.\par

In contrast, quantum estimation theory provides a mathematical framework for optimizing measurements in quantum systems. It is particularly relevant when estimating a parameter through repeated measurements on identically prepared quantum systems \cite{Paris2009,B4}. To explore how quantum properties influence measurement precision, the dynamics of a system are modeled using quantum Fisher information. Recent research has revealed deep connections between quantum Fisher information and various quantum phenomena, such as quantum phase transitions \cite{Ye2016}, quantum correlations \cite{Kim2018,B7}, quantum thermodynamics \cite{Hasegawa2020,Hminat}, entanglement manipulation \cite{Chapeau2017}, and the quantum speed limit \cite{Taddei2013,B6}.\par

Quantum thermometry, an emerging subfield of quantum metrology, focuses on determining the temperature of quantum systems, offering exciting possibilities for indirect measurement techniques \cite{Michalski2002}. This discipline holds significant potential for achieving high-precision measurements at low temperatures \cite{DePasquale2016} through the use of quantum thermometers \cite{Scigliuzzo2020}. An optimal quantum thermometer is typically required to be significantly smaller than the system it measures \cite{Brunelli2012}.\par

In quantum mechanics, temperature does not constitute a quantum observable. While it retains its thermodynamic significance, it lacks a direct operational definition in a quantum context. 
The selection of an estimator involves classical post-processing of measurement outcomes  \cite{B8}, while the choice of measurement itself is a core challenge in quantum metrology \cite{B5}, as different measurements yield varying levels of precision. Quantum parameter estimation has been effectively utilized in diverse areas, including phase estimation \cite{30}, dynamics of open systems and quantum measurements in qubit arrays \cite{36} , and quantum phase transitions \cite{39}. It is important to clarify that we are not addressing temperature fluctuations in a thermodynamic sense \cite{B1,B2,B3}. However, the estimated temperature value, derived from measurements on the qubit, is subject to fluctuations \cite{41}. In our analysis of the interaction model, we consider the qubit's coupling with the sample to be described by a dephasing Hamiltonian. This framework allows for an analytical solution \cite{43} when the environment, represented by the sample, is in a thermal state. For quantitative evaluation, we employ an Ohmic spectral density \cite{45} characterized by a general Ohmicity parameter \cite{48}. The resulting dynamics stem from a complex interplay between the dephasing mechanism and the Ohmic properties of the environment \cite{51}, particularly pronounced at low temperatures. Previous studies have shown that parity-time -symmetric qubits exhibit superior decoherence characteristics compared to Hermitian qubits when weakly coupled to a Hermitian environment \cite{3}. Weak coupling to the bath ensures minimal heat exchange between the system and the environment \cite{4}, leading to a process termed pure decoherence or dephasing \cite{5}. We adopt a Hermitian bath in this study.\par

To investigate these properties, we first analyze the dynamics of the qubit systems. The reduced density matrix of the qubit, derived from the expectation value of the time-evolved operator, can be readily calculated for both Hermitian and PT-symmetric qubits using a Hermitian diagonal qubit \cite{6}. In contrast, for APT qubits, we present a method for deriving the exact analytical form of the time-dependent density matrix for a generic non-Hermitian qubit system. This is accomplished by utilizing a time-dependent Dyson map to identify an equivalent Hermitian system, thereby deriving the reduced density matrix for a time-dependent Hermitian system \cite{7}. Following the development of parity-time  symmetry \cite{9} and two decades of comprehensive research \cite{10}, the concept of anti-parity-time  symmetry was introduced by Ge and Türeci \cite{11} within the framework of optics, utilizing a specific configuration. For the effective optical potential, the $\mathcal{PT}$ operator commutes with the Hamiltonian, satisfying $[H, \mathcal{PT}] = 0$, whereas the anti-$\mathcal{PT}$ operator anticommutes with it, satisfying $\{ H, \mathcal{PT} \} = 0$, whereas the anti-PT operator anticommutes with it, satisfying $\{H, PT\} = 0$. Experimental demonstrations of anti-$\mathcal{PT}$ symmetry have been achieved in multiple physical settings, such as spatially coupled atomic beams~\cite{12}, resonators in electrical circuits~\cite{13}, optical waveguides with purely imaginary coupling constants~\cite{14}, and optical four-wave mixing processes in cold atomic media~\cite{15}.
Furthermore, constant-refraction optical systems \cite{16}, as well as experiments in atomic \cite{17} and optical \cite{19} setups, have demonstrated anti-PT symmetry. Its applications span diverse areas such as waveguide arrays \cite{21}, diffusive systems \cite{22}, phase transitions \cite{23}, spin chains \cite{24}, information flow \cite{25}, and non-Markovian dynamics \cite{26}.\par

In this study, we focus on an anti-PT-symmetric quantum system weakly coupled to an environment (or bath). This setup significantly reduces decoherence rates, outperforming PT-symmetric systems. It is conceivable that an anti-PT qubit could be implemented in recently developed optical and microcavity systems \cite{27}. Recent studies have also investigated quantum circuits \cite{28} and information flow \cite{29} in two-level systems. The paper is structured as follows. Section \ref{Sec2} offers a brief overview of local quantum estimation theory  tools. Section \ref{sec3} describes our physical model and the application of QET methods. Section \ref{sec4} presents our results, illustrating how  anti-PT symmetry provides the best protection against decoherence and the highest QFI, highlighting the impact of environmental structure at low temperatures and the advantage of non-Hermitian symmetries for quantum thermometry. Section \ref{sec5} demonstrates that the QSNR remains low at low temperatures and increases with T, and saturates to a universal value at high T; it is highest under Hermitian symmetry and in the sub-Ohmic regime, favoring precise temperature estimation. and Section \ref{sec6} provides concluding remarks.

\section{Apparatus of Quantum Estimation Theory}\label{Sec2}

In numerous sensing methodologies, direct access to the physical parameters of interest is unavailable, necessitating their assessment through indirect measurements. Consequently, the objective of estimation theory is to deduce the value of the target quantity by analyzing data derived from the measurement of an alternative observable. We explore the optimal estimation of the temperature \( T \) of a bosonic thermal bath by conducting measurements on a quantum probe allowed to interact with the bath, which serves as its environment. The state of the probe is characterized by a density operator \( \rho \), which through interaction with the environment, evolves into a temperature-dependent form, \( \rho \to \rho_T \).

We represent the conditional probability distribution of the measurement outcomes of \( X \), given the true parameter value 
 \( T \), as \( p(x|T) \). An estimator, denoted \( \hat{T}(x) \), where 
 \begin{equation}
 x = \{x_1, x_2, \dots, x_M\}, 
 \end{equation} 
 is defined as any function that maps the \( M \) observed measurement outcomes to an estimated value of the parameter of interest. The expectation value of this estimator is determined by the probabilistic outcomes of the measurements, providing a statistical prediction of the parameter. The expected value of the estimator is given by
\begin{equation}
\overline{T} = \int dx \, p(x|T) \, \hat{T}(x),
\end{equation}
while the precision of the estimation process is characterized by the variance of the estimator, expressed as
\begin{equation}
\text{Var} \, \hat{T} = \int dx \, p(x|T) \, \left[ \hat{T}(x) - \overline{T} \right]^2.
\end{equation}
In these expressions, the probability distribution satisfies \( p(x|T) = \prod_{k=1}^M p(x_k|T) \), as the measurements are conducted on independently prepared instances of the system. For any unbiased estimator (i.e., one where \( \overline{T} \to T \) in the limit of large \( M \)), the variance is constrained by the Cramér-Rao bound, which is articulated as
\begin{equation}
\text{Var} \, \hat{T} \geq \frac{1}{M F(T)},
\end{equation}
where \( F(T) \) denotes the Fisher information associated with the measurement of \( X \), defined by
\begin{equation}
F(T) = \int dx \, p(x|T) \, \left[ \partial_T \ln p(x|T) \right]^2.
\end{equation}

The most suitable measurement for estimating the parameter $T$ is the one that yields the maximum Fisher information. 
An estimator is considered efficient when it reaches equality in the Cramér–Rao bound. 
When such an efficient estimator is combined with the optimal measurement, it constitutes an ideal strategy for determining $T$. 
Maximizing over all possible quantum measurements leads to the Quantum Fisher Information (QFI), denoted by $H(T)$. 
The ultimate precision permitted by quantum mechanics is achieved when the estimator attains the quantum Cramér–Rao limit
\begin{equation}
\text{Var} \, \hat{T} \geq \frac{1}{M H(T)}.
\end{equation}
The QFI can be derived from the quantum state of the system, specifically from its eigenvalues and eigenvectors, which encapsulate the dependence on the parameter. For a system described by a density matrix in its diagonal form,
\begin{equation}
\rho_T = \sum_n \rho_n |\phi_n\rangle\langle\phi_n|,
\end{equation}

the QFI is expressed as
\begin{equation}
H(T) = \sum_p \frac{(\partial_T \rho_p)^2}{\rho_p} + 2 \sum_{n \neq m} \frac{(\rho_n - \rho_m)^2}{\rho_n + \rho_m} |\langle \phi_n | \partial_T \phi_m \rangle|^2.
\end{equation}
The first term in this expression reflects the parameter dependence of the eigenvalues and is known as the classical contribution to the QFI, while the second term, referred to as the quantum contribution, accounts for the parameter dependence of the eigenvectors. It is worth noting that the local quantum estimation theory (QET) described here, which depends on the specific value of \( T \), presupposes some prior approximate knowledge of the parameter.\par

To evaluate the estimability of a parameter without relying on its specific value, the signal-to-noise ratio (SNR) , defined as 
\begin{equation} R_T = T^2 / \text{Var} \, \hat{T},
\end{equation}
where higher values indicate superior estimators. By applying the quantum Cramér-Rao bound, we obtain the inequality
\begin{equation}
R_T \leq Q_T \equiv T^2 H(T),
\end{equation}
where \( Q_T \) is known as the quantum signal-to-noise ratio (QSNR). A larger QSNR signifies that the parameter \( T \) can be estimated more effectively. In the following analysis, we employ this framework to determine the temperature $T$ of a structured sample characterized by an Ohmic spectral density, employing a qubit that interacts with the sample over a fixed time interval, after which a measurement is performed to deduce the temperature. Specifically we investigate the feasibility of an effective quantum thermometric protocol using an exactly solvable qubit dephasing interaction model. To achieve this, we identify the optimal interaction time that maximizes the QFI and compute the associated Quantum Signal-to-Noise Ratio. Our results include encouraging numerical findings across all scenarios examined, alongside analytical expressions derived for the super-Ohmic regime, the low-temperature limit in an Ohmic environment, and the high-temperature case across different spectral density profiles.

\section{Physical model}\label{sec3}

We begin our analysis by considering a quantum probe composed of a single qubit with energy level separation denoted by \(\omega_0\). We consider a spin qubit system interacting with a bath of bosonic modes. We adopt natural units by setting \(\hbar = 1\) and normalize all frequency scales with respect to \(\omega_0\). The resulting total dimensionless Hamiltonian of the system can be expressed as  \begin{equation}
H = H_S \otimes I_B + I_S \otimes H_B + H_S \otimes H_I \end{equation}  where \( H_S \) represents the system Hamiltonian, \( H_B \) describes the bath, given by
\begin{equation}
H_B = \sum_k \omega_k b^\dagger_k b_k,
\end{equation}
and \( H_I \) denotes the interaction term between the system and the bath, expressed as
\begin{equation}
H_I = \sum_k \left( g_k b^\dagger_k + g^*_k b_k \right).
\end{equation}
The operators \( b_k^\dagger \) and \( b_k \) represent the bosonic creation and annihilation operators corresponding to the \( k \)-th mode of the environment. These operators satisfy the canonical commutation relation \( [b_k, b_{k'}^\dagger] = \delta_{k k'} \). The term \( \omega_k \) denotes the frequency of the \( k \)-th mode of the bath, while \( g_k \) indicates the interaction strength between the qubit and the corresponding mode. Both \( \omega_k \) and \( g_k \) are expressed in units of \( \omega_0 \), rendering them dimensionless quantities. A primary objective of this study is to analyze and compare the characteristics and behaviors of Hermitian and non-Hermitian qubit systems:\par

{\bf ($i$) Hermitian Symmetric Qubit:} We begin by examining the dynamics of the Hermitian qubit, characterized by the Hamiltonian
\begin{equation}
H_H^S = \begin{pmatrix}
a + d & c + i b \\
c - i b & -a + d
\end{pmatrix},
\end{equation}

where \(a, b, c, d \in \mathbb{R}\). To satisfy the Hermiticity condition $H_H^S = (H_H^S)^\dagger$. For convenience, we take a similarity transformation~\cite{33}
\[
T = \begin{pmatrix}
\omega_0 - a & - c - i b \\
\omega_0 + a & c + i b
\end{pmatrix}, \quad \text{where} \quad \omega_0 = \sqrt{a^2 + b^2 + c^2}.
\]
We diagonalize the Hamiltonian by applying the similarity transform \(H_{Dh}^S = T H_h^S T^{-1}\). The system’s time-evolved reduced density matrix can then be written in terms of the reduced density matrix in the diagonal representation \cite{3b}, with normalization given by
\begin{equation}
\rho_S(t) = T^{-1} \rho_{Dh}^S(t) (T^{-1})^\dagger \frac{1}{\text{Tr}_S \left[ T^{-1} \rho_{Dh}^S(0) (T^{-1})^\dagger \right]}.
\end{equation}

{\bf ($ii$) PT-symmetric  Qubit (PT):} For the PT-symmetric case, the Hamiltonian is given by
\begin{equation}
H_{PT}^S = \begin{pmatrix}
a + i d & c + i b \\
c - i b & a - id
\end{pmatrix},
\end{equation}
and the constraint \(b^2 + c^2 \geq d^2\). The parity and time operators are defined as
\[
P = \sigma_x = \begin{pmatrix}
0 & 1 \\
1 & 0
\end{pmatrix}, \quad T: i \to -i,
\]
satisfying the commutation relation \([PT, H_{PT}^S] = 0\) , similarity to the Hermitian system , but with
\begin{equation}
\omega_0 = \sqrt{a^2 + b^2 - d^2},
\end{equation}
yields a diagonalized Hermitian Hamiltonian \(h_{PT}^S = T H_{PT}^S T^{-1}\), and the evolved reduced density matrix , with the distinction that \(\Omega(t)\) is replaced by \(-\Omega(t)\).\par

{\bf ($iii$) Anti-PT-symmetric  Qubit (APT):} Next, we introduce an APT symmetric quantum system, described by the general Hamiltonian
\begin{equation}
H^S_{APT} = \begin{pmatrix}
a + i d & c + i b \\
- c + i b & - a + i d
\end{pmatrix},
\end{equation}
 One finds \(\{PT, H^S\}=0\) with \(P\) (parity) and \(T\) (time reversal). Therefore the complete Hamiltonian \(H\) may be brought to diagonal form via the same similarity mapping previously employed;
\begin{equation}
\omega_0 = \sqrt{a^2 - b^2 - c^2}.
\end{equation}

The resulting diagonalized Hamiltonian is expressed as
\[
H_D = T H T^{-1}.
\]

This is equivalent to the eigenbasis representation
\begin{equation}
T H T^{-1} = \sum_n E_n |n\rangle\langle n|,
\end{equation}
where \(E_n \in \mathbb{C}\). This can be rewritten in a complete biorthonormal basis for \(H\)~\cite{37}
\begin{equation}
H= \sum_n E_n |\psi_n^R\rangle\langle \psi_n^L|,
\end{equation}
where the set of eigenvectors is defined as \(|\psi_n^R\rangle = T^{-1} |n\rangle\) and \(\langle \psi_n^L| = \langle n| T\), and satisfying:
\begin{equation}
\langle \psi_n^L | \psi_m^R \rangle = \delta_{nm} \quad \text{and} \quad \sum_n |\psi_n^R\rangle\langle \psi_n^L| = I.
\end{equation}

The eigenvalues of the APT symmetric system are expressed as \( E_\pm = i d \pm \omega_0 \), with the energy difference between them given by \( E_e = 2 \omega_0 \). This quantity represents the energy gap of the qubit system, and its real nature depends on the parameters \( a \), \( b \), and \( c \). When the condition \( b^2 + c^2 = a^2 \) is satisfied, the energy gap vanishes, leading to degenerate eigenvalues. However, for the purposes of this paper, we focus on the case where the energy gap is real,the parametric domain \(a^2 \geq b^2 + c^2\) . The combined qubit parameter \(\omega_0\) is defined as
\begin{equation}
\omega_0 = \begin{cases} 
\sqrt{a^2 + b^2 + c^2} & \text{(Hermitian)} \\
\sqrt{a^2 + b^2 - d^2} & \text{(PT)} \\
\sqrt{a^2 - b^2 - c^2} & \text{(APT)}
\end{cases},
\end{equation}

In the remainder of the paper we will take  $c = d$ and the time-dependent phase function \(\Omega(\tilde{t})\) is given by
\begin{equation}
\Omega(\tilde{t}) = \begin{cases} 
\Omega(t) & \text{(Hermitian)} \\
-\Omega(t) & \text{(PT)} \\
\Omega_2(t) - \Omega_1(t) & \text{(APT)}
\end{cases},
\end{equation}
where the expressions for the functions \(\Omega(t)\), \(\Omega_1(t)\), and \(\Omega_2(t)\) are given as follows (for more details consult \cite{3b}):
\begin{equation}
\Omega(t) = 4\theta \int_{0}^{\infty} dw \, J(w) \frac{wt - \sin(wt)}{w},
\end{equation}
\begin{equation}
\Omega_1(t) = 4\theta \int_{0}^{\infty} dw \, J(w) \frac{1 - \cos(wt)}{w^2}, 
\end{equation}
\begin{equation}
\Omega_2(t) = 2\theta t^2 \int_{0}^{\infty} dw  J(w), 
\end{equation}
in the continuum limit of bath modes. Here, the spectral density characterizing the environment and which we will address in more depth in the next section. In general, a complex coupling constant \(g_k = |g_k| e^{i \theta_k}\).
Upon transformation to the interaction picture, the non-unitary evolution of the probe, resulting from its coupling to the thermal bath, is described by a reduced dynamical map acting on the system’s state. For a general non-Hermitian Hamiltonian \(H\), it can be decomposed into real and imaginary components, \(H = H_R + i H_I\), leading to a complex extension of the Liouville-von Neumann equation:
\begin{equation}
\dot{\rho}_t = -i [H_R, \rho] + \{H_I, \rho\} - 2\rho \text{Tr}(\rho H_I).
\end{equation}
This equation can be solved using the form
\begin{equation}
\rho(t) = \frac{U(t) \rho(0) U^\dagger(t)}{\text{Tr}[U(t) \rho(0) U^\dagger(t)]}.
\end{equation}
\begin{figure*}
        \includegraphics[width=5.9cm]{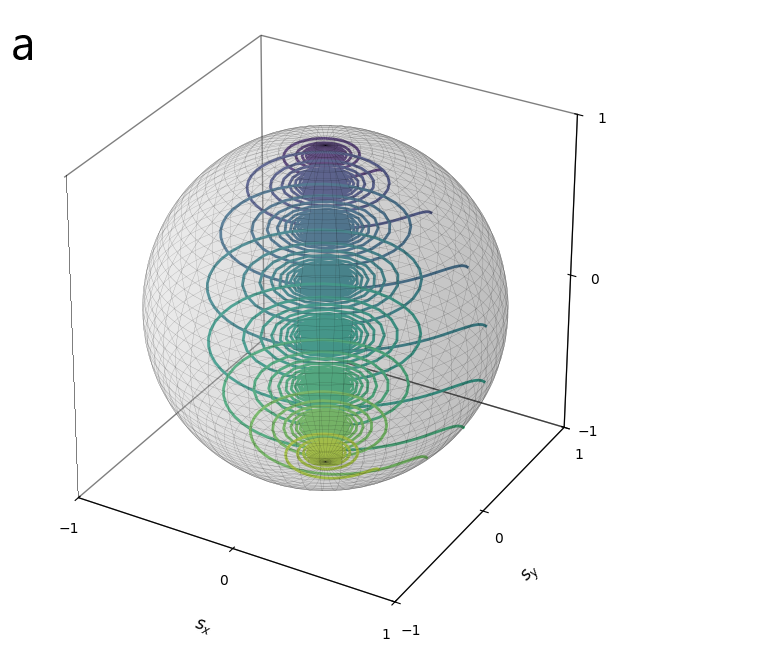}
        \includegraphics[width=5.9cm]{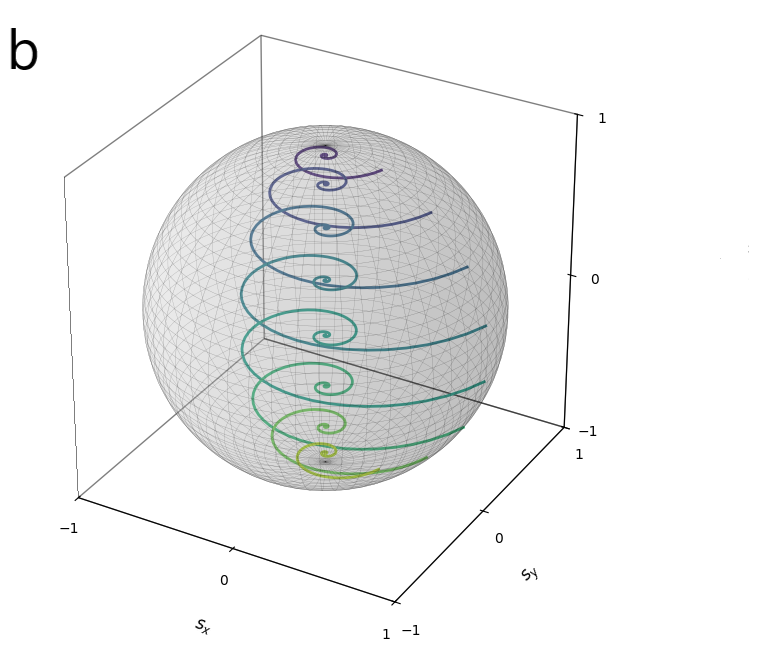}
        \includegraphics[width=5.9cm]{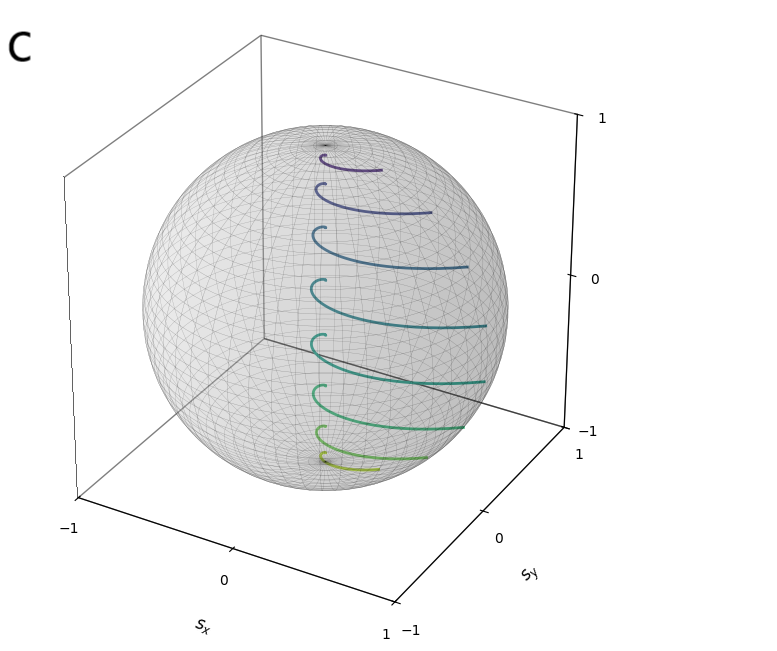}
      \caption{Bloch sphere trajectories of Hermitian (top left), PT-symmetric (top right), and APT-symmetric (bottom) qubits, showing their evolution for various initial \(\theta_0\) values, using parameters \(J_0 = 1\), \(\omega_c = 1\), \(s = 0.5\), \(T = 2.0\), \(b = 0.6\), \(c = 0.6\), \(d = 0.6\), and \(a = 1.0\).}
    \label{fig2}
\end{figure*}
The quantum state trajectories in Fig.(\ref{fig2}) commence at the periphery of the Bloch sphere and proceed to spiral inward toward its center. Initially, all three examined cases exhibit a clockwise rotational evolution. However, both the Hermitian and APT symmetric systems transition to an anticlockwise motion shortly thereafter. This shift is particularly pronounced in the APT symmetric case, where the change in direction is readily discernible. In contrast, the Hermitian case presents a subtler transition, as the rapid convergence of the trajectory toward the sphere's center outpaces the temporal scale of the directional change.

\section{Quantum thermometry by dephasing} \label{sec4}
\subsection{ Quantum dephasing process and optimal preparation} 
\begin{figure}
    \centering
    \includegraphics[width=1.0\linewidth]{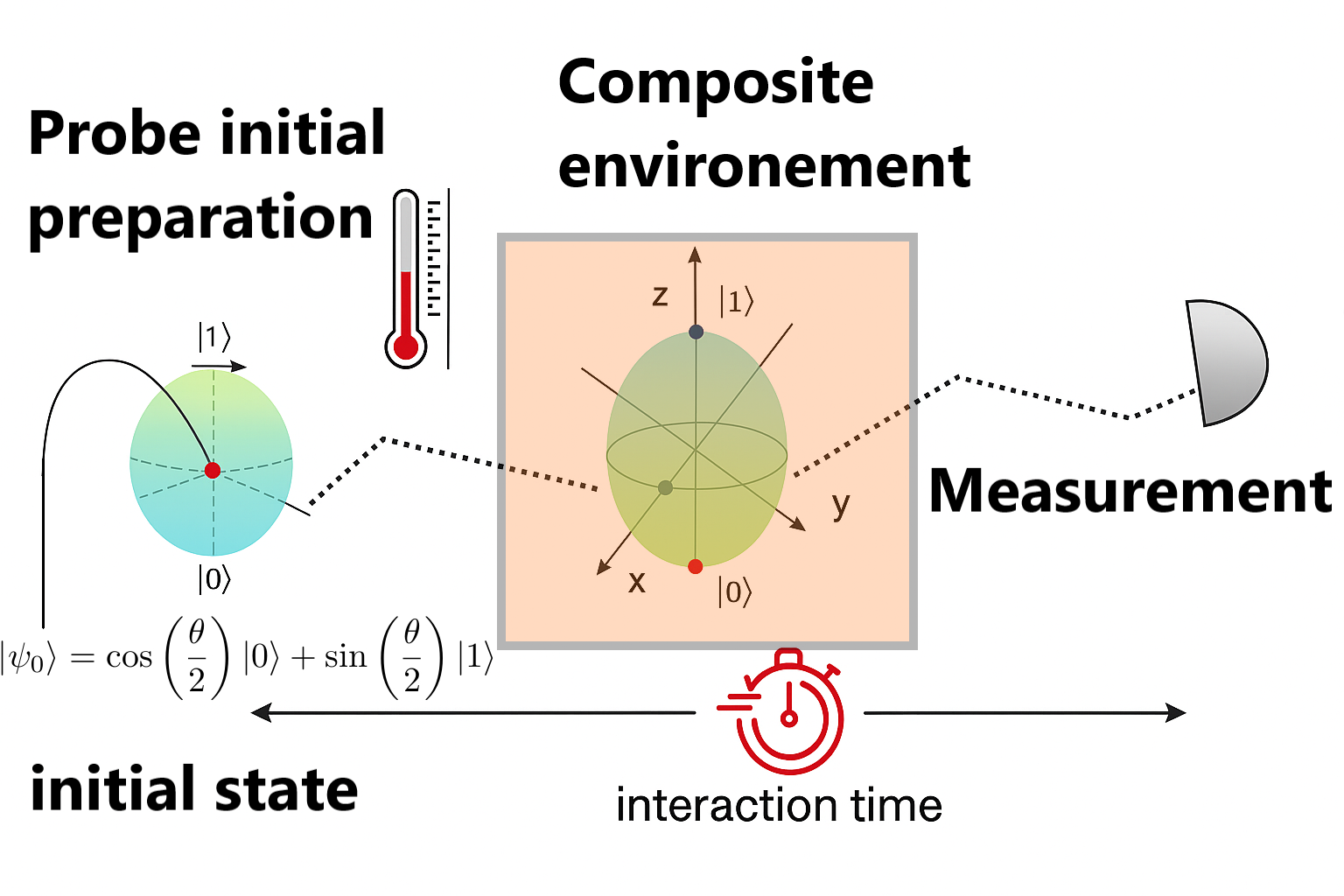} 
    \caption{A quantum thermometric protocol is proposed, utilizing a single qubit system subjected to dephasing induced by its coupling to a structured thermal reservoir in equilibrium. The dephasing process manifests as a contraction of the Bloch sphere within the interaction picture. Following the interaction, a measurement is conducted along the optimal spin orientation to probe the system's state.}
    \label{fig1}
\end{figure}

Any coupling between a quantum system and its surrounding environment typically alters the relative phases of the components of its wavefunction. This phenomenon leads to dephasing commonly followed by decoherence arising from interactions between the system and the various modes of the thermal reservoir. Interestingly, this dephasing mechanism in Fig.(\ref{fig1}) can be leveraged to employ the quantum system as a sensitive probe for estimating environmental parameters, all while leaving the system’s energy content essentially unaffected.
The  environment is modeled as a thermal reservoir in equilibrium at temperature \( T \), while initial state of the qubit probe is assumed to be in a state of the form 
\begin{equation}
|\psi_0\rangle = \cos\left(\frac{\theta}{2}\right) |0\rangle + \sin\left(\frac{\theta}{2}\right) |1\rangle,
\end{equation}

The QFI can be analytically computed  after diagonalizing the density matrix for the qubit \(\rho_0 = |\psi_0\rangle\langle\psi_0|\) . The evolved density matrix \(\rho(t)\) is obtained by computing the element-wise Hadamard product denoted by where \(\circ\);
\begin{equation}
\rho(t) = \Phi(t) \circ \rho(0) .
\end{equation}
The dynamical map $\Phi(t)$ is given by 
\begin{equation}
\Phi(t) = \begin{pmatrix}
1 & e^{-\Gamma(t, T)-i\varphi(t)} \\
e^{-\Gamma(t, T)+i\varphi(t)} & 1
\end{pmatrix}.
\end{equation}

Here, \(\Gamma(t,T)\) is the decoherence factor combined with the  imaginary phase \(\phi(t)\), dependent on the bath’s spectral density, and \(\omega_c\) is the phase will exert no significant influence on the dynamics of our system or its quantum resources. Henceforth, we shall consider the phase to be constant for reasons of convenience. The time evolution of the off-diagonal elements is governed by the decoherence factor
\begin{equation}
e^{-\Gamma(T, t)} = \sum_k \left\langle \exp(g_k b_k^\dagger - g_k^* b_k) \right\rangle  \end{equation}
where \(\left\langle \cdot \right\rangle = \text{Tr}[\cdot \rho_B]\) denotes the expectation value over the thermal state of the bath \(\rho_B\). The decoherence factor \(\Gamma(T, t)\) is influenced by the temperature and the coupling frequency distribution of the bath, characterized by the spectral density \(J(\omega)\). In this study, we model the environment using a reservoir characterized by a spectral density function that falls within the Ohmic class. The spectral density is given by  
\begin{equation}
J(\omega, \omega_c) =J_0  \frac{ \omega^s}{ \omega_c^{s - 1}} e^{-\omega / \omega_c},
\end{equation}
where \(s\) is a positive real parameter that classifies the environment , specifically, we explore three distinct values of the Ohmicity parameter, corresponding to representative cases of sub-Ohmic ($s=0.5$), Ohmic ($s=1$), and super-Ohmic ($s=3$) regimes. The parameter \(\omega_c\) denotes the cutoff frequency (we set \(\omega_c\) = 1) .
 Consequently, the initial density matrix for the composite system is expressed as:

\begin{equation}
\rho(t) = \begin{pmatrix}
\cos^2 \frac{\theta}{2}                    & \sin \theta e^{-i\varphi(t)} e^{-\omega_0^2 \gamma(t)} \\
\sin \theta e^{i\varphi(t)} e^{-\omega_0^2 \gamma(t)} & \sin^2 \frac{\theta}{2}
\end{pmatrix}, 
\end{equation}
where the phase is defined by
\begin{equation}
\varphi(t) = \varphi_0 - 2\omega_0 t + \omega_0 \Omega(\tilde{t}).
\end{equation}
Giving:\begin{equation}
H(T, t) = \sin^2 \theta  \frac{[\partial_T \Gamma(T, t)]^2}{e^{2 \Gamma(T, t)} - 1}.
\end{equation}
\begin{widetext}
\begin{figure*}[t] 
      
\includegraphics[width=5.9cm]{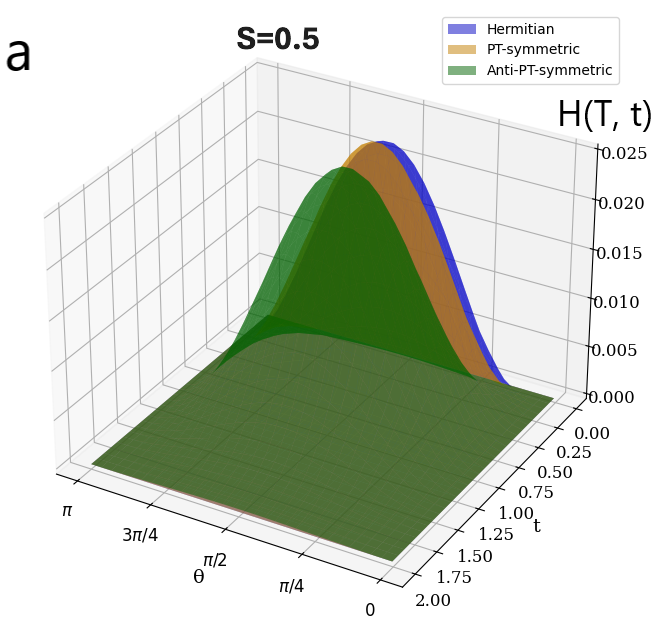}
        \includegraphics[width=5.9cm]{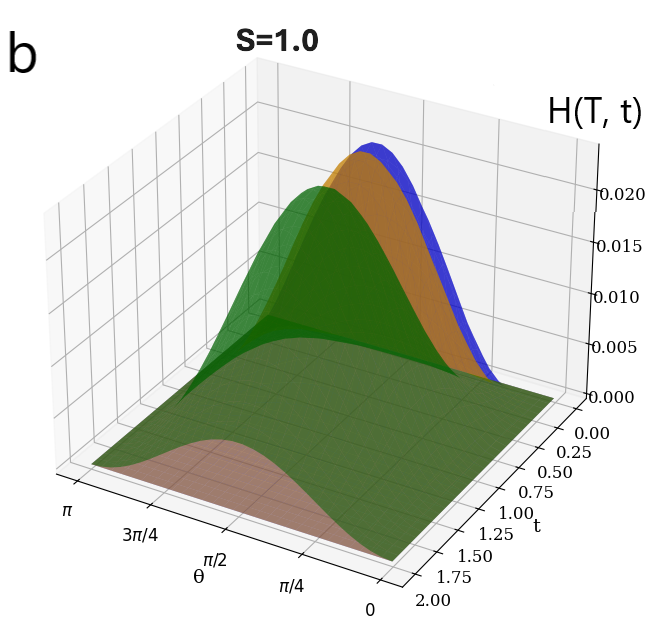}
        \includegraphics[width=5.9cm]{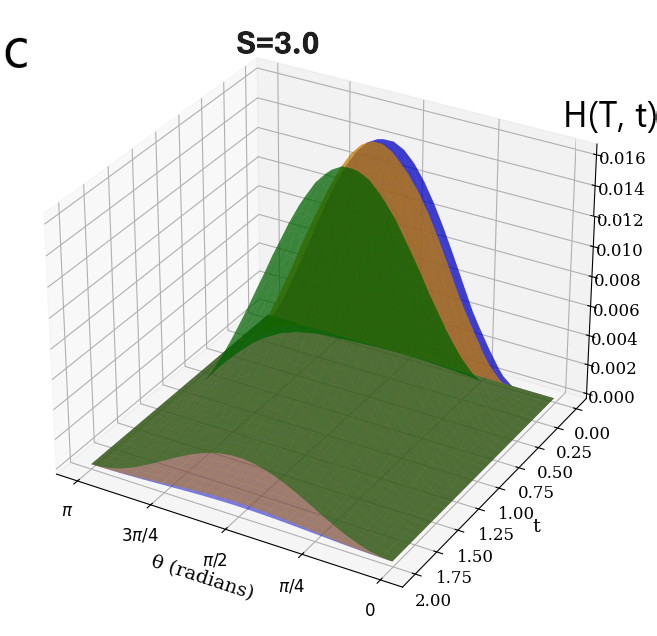}
      \caption{We show the QFI $H(T, t)$ as a function of both the angle of state preparation $\theta$ and the interaction time $t$ for three illustrative types of structured environments: (a) Ohmic ($s = 1$), (b) sub-Ohmic ($s = 0.5$), and (c) super-Ohmic ($s = 3$). Each case is analyzed under different symmetry conditions, including both Hermitian and non-Hermitian dynamics.}
    \label{fig3}
\end{figure*}
\end{widetext}

After transforming to the interaction picture, the non-unitary dynamics of the probe, arising from its interaction with the thermal bath, are captured by a reduced dynamical map acting on the system's state. Notably, this optimal configuration remains unaffected by both the temperature \( T \) and the duration of the interaction. The QFI is maximized when \( \theta = \frac{\pi}{2} \), leading to the optimal initial state preparation \( |+\rangle = \frac{1}{\sqrt{2}} (|0\rangle + |1\rangle) \), which is independent of \( T \), the interaction time, and the Ohmicity parameter \( s \) of the reservoir (see Fig.~\ref{fig3} for details).\par

Inserting \(\Gamma(t,\omega_c)\) as given in Eq.~(6) into the preceding equation yields an analytical formula for the decoherence coefficient for a specified value of \(s\). To optimize the estimation, we search for the interaction time \(t\) that maximizes the QFI versus temperature \(T\) (cf. Fig.~\ref{fig1}), with \(s\) held fixed . The maximization of the QFI over time has been performed numerically. The full decoherence factor, incorporating thermal effects, is:
\begin{multline}
\Gamma(T, t) = \Gamma(0, t) + 2 \sum_{n=1}^\infty \left( 1 + \frac{n}{T} \right)^{1-s} \\
\times \Gamma \left( 0, \frac{t}{1 + \frac{n}{T}} \right),
\end{multline}
where:
\begin{equation}
\Gamma \left( 0, \frac{t}{a_n}\right) = \begin{cases}
\frac{1}{2} \log \left( 1 + \left( \frac{t}{a_n} \right)^2 \right) & s = 1, \\
\begin{aligned}
&\Gamma(s-1) \left[ 1 - \left( 1 + \left( \frac{t}{a_n} \right)^2 \right)^{\frac{1-s}{2}} \right. \\
&\quad \left. \times \cos \left[ (1-s) \arctan \left( \frac{t}{a_n} \right) \right] \right]
\end{aligned} & s \neq 1
\end{cases}
\end{equation}

with $a_n$ = 1 + $\frac{n}{T}$.

\subsection{Quantum thermometry for different Symmetries} 
\begin{figure*}[t] 
        \includegraphics[width=5.8cm]{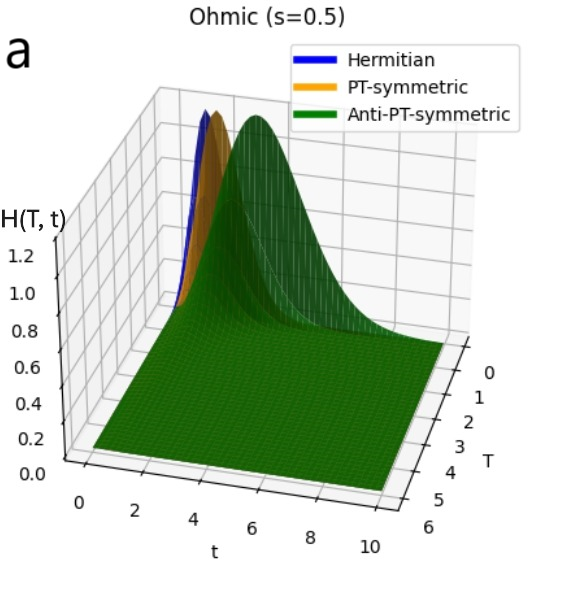}
        \includegraphics[width=5.8cm]{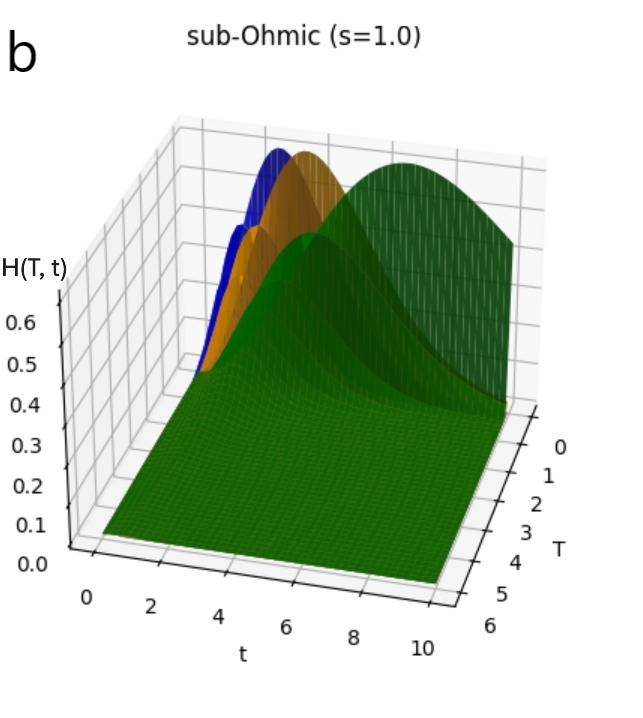}
        \includegraphics[width=5.8cm]{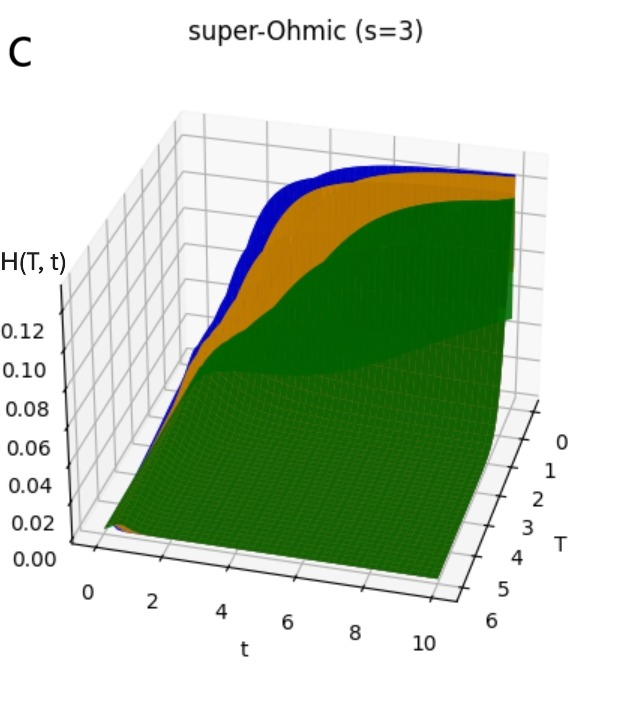}
      \caption{We show the QFI $H(T, t)$ as a function of both the reservoir temperature $T$ and the interaction time $t$ for three illustrative types of structured environments: (a) Ohmic ($s = 1$), (b) sub-Ohmic ($s = 0.5$), and (c) super-Ohmic ($s = 3$). Each case is analyzed under different symmetry conditions, including both Hermitian and non-Hermitian dynamics.}
    \label{fig4}
\end{figure*}

In this study, we investigate quantum thermometry across various samples characterized by spectral densities within the Ohmic class.  The behavior of the QFI  as a three-dimensional representation of \( H(T, t) \), as a function of temperature \( T \) and time \( t \) in Fig.\ref{fig4}.(a-c) across different symmetry conditions  (hermitian and non hermitian). A qualitatively similar behavior is observed across all three symmetry regimes. However, a notable difference lies in the rate at which information diffuses from the system into the environment, where the varying slopes observed in these graphs delineate distinct thermal regimes, specifically those corresponding to low and high temperatures. At lower temperatures, the optimal interaction time is extended, whereas it diminishes progressively as the temperature increases. This trend aligns with expectations based on the nature of the probing mechanism. The probe's capacity to acquire information regarding the environmental temperature stems from its susceptibility to decoherence. At low temperatures, decoherence is less pronounced, requiring a longer duration to encode information onto the probe. Conversely, at elevated temperatures, decoherence occurs more rapidly. Notably, at high temperatures, decoherence is predominantly driven by thermal fluctuations specifically for hermitian symmetry, rendering the environmental structure largely irrelevant. This is evidenced by the consistent behavior of the QFI across the three values of the Ohmicity parameter at high temperatures. In contrast, at low temperatures, the environmental structure plays a critical role in shaping the decoherence mechanism, leading to a pronounced dependence of the QFI on the Ohmicity parameter. The presence of a finite QFI maximum indicates that optimal temperature and time estimation can be achieved prior to the qubit reaching its steady state at a finite interaction time $t_{\text{opt}}$ and temperature $T_{\text{opt}}$, the symmetry does not affect the overall behavior but rather the rate at which decoherence occurs.

This study underscores the nuanced interplay between symmetry, spectral density, and decoherence dynamics in quantum thermometry within Ohmic-class environments. The consistent superiority of anti-PT symmetry in yielding the highest QFI across all regimes highlights its enhanced resilience against decoherence, offering a promising avenue for precise temperature estimation. Conversely, Hermitian dynamics exhibit greater susceptibility to environmental noise, resulting in the lowest QFI, while the noise spectrum’s influence reveals stronger sensitivity in Ohmic and sub-Ohmic regimes. These findings advocate for leveraging non-Hermitian symmetries to optimize quantum sensing protocols under diverse thermal conditions.

\subsection{Thermal optimization} 
In this subsection, we will explore the pathways for temporal and thermal optimization to maximize the QFI under different symmetry conditions.

\begin{figure*}[t] 
        \includegraphics[width=5.9cm]{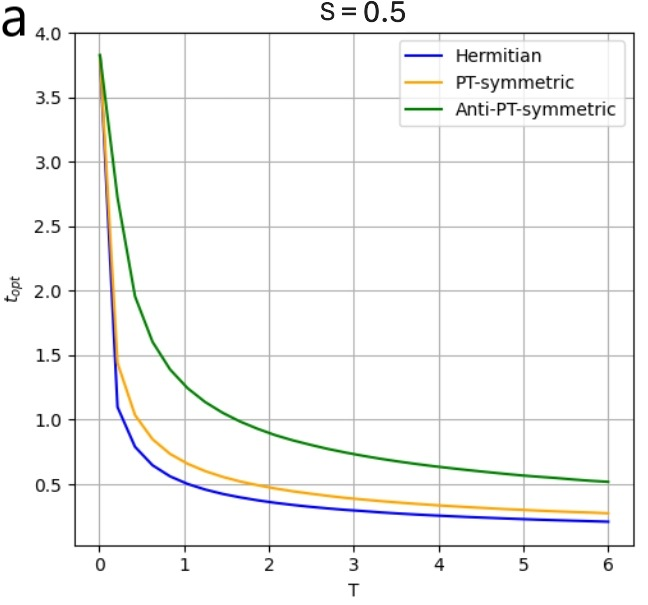}
        \includegraphics[width=5.9cm]{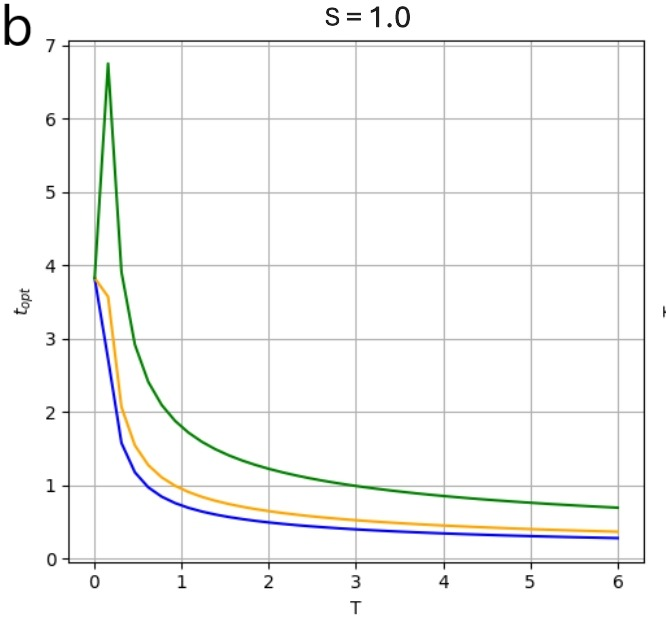}
        \includegraphics[width=5.9cm]{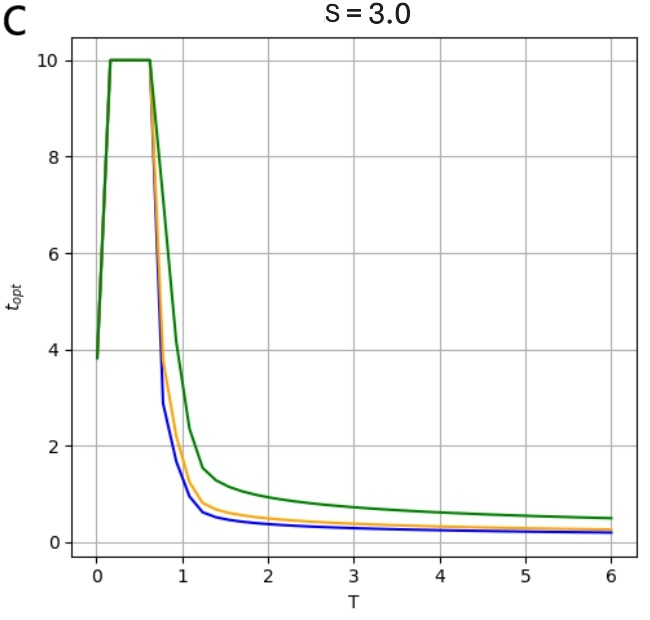}
        \includegraphics[width=5.9cm]{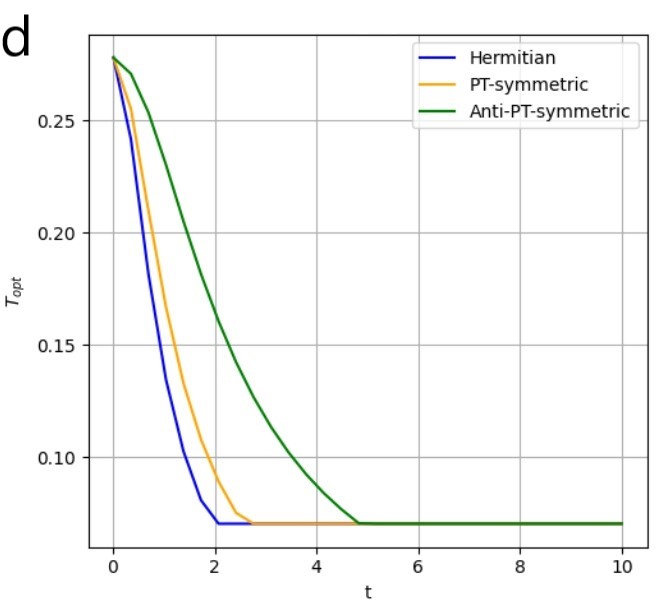}
        \includegraphics[width=5.9cm]{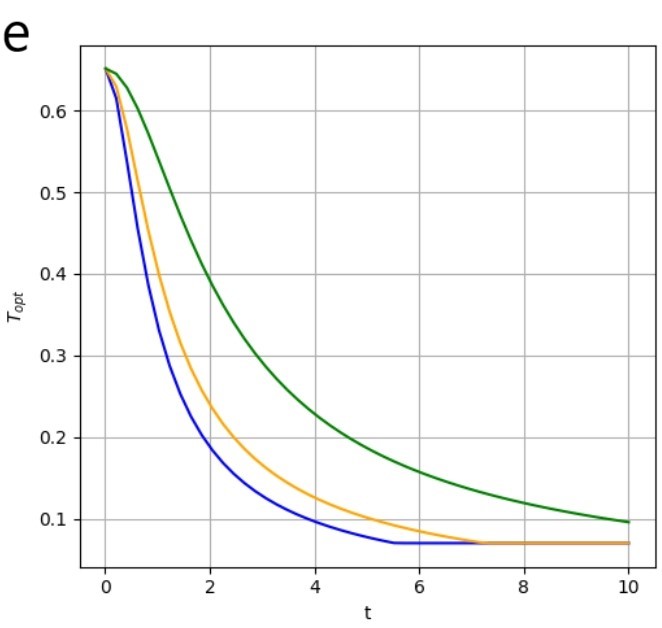}
        \includegraphics[width=5.9cm]{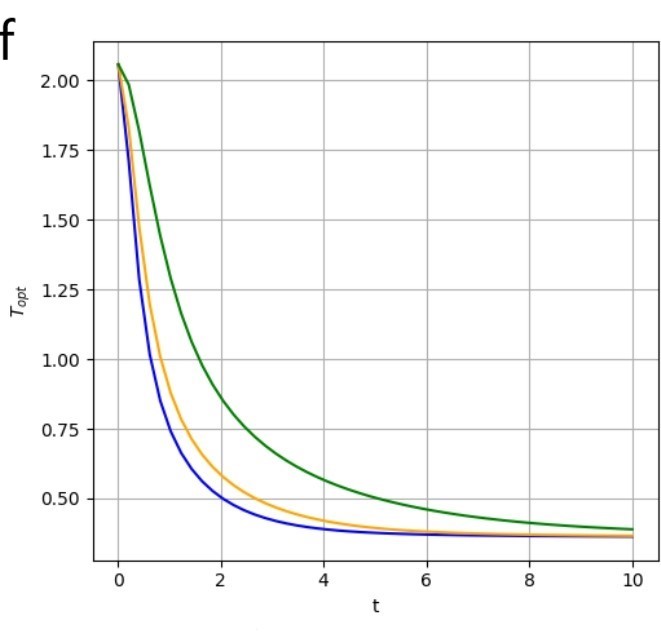}
      \caption{We report the optimal interaction times $t_{\text{opt}}$ as a function of the reservoir temperature $T$. Specifically, we plot the temperature $T_{\text{opt}}$ at which the QFI reaches its maximum, as a function of the interaction time $t$. Three characteristic types of structured environments have been analyzed: (a) Ohmic reservoir ($s = 1$), (b) sub-Ohmic reservoir ($s = 0.5$), and (c) super-Ohmic reservoir ($s = 3$), under both Hermitian and non-Hermitian dynamics.}
    \label{fig5}
\end{figure*}

The  Fig.\ref{fig5}.(a-c) present dual plots for the Ohmicity parameters \( s = 0.5 \), \( s = 1.0 \), and \( s = 3.0 \), showcasing the optimal interaction time \( t_{\text{opt}} \) under Hermitian (blue), PT-symmetric (orange), and anti-PT-symmetric (APT, green) symmetries, which critically influence QFI dynamics for temperature estimation in a quantum probe coupled to a reservoir. The upper plots illustrate \( t_{\text{opt}} \), the time at which the QFI peaks for a fixed temperature, revealing a decline from initial values, with APT consistently exhibiting the slowest decay due to non-Hermitian effects near exceptional points that suppress environmental coupling, thereby preserving coherence longer. PT-symmetric systems, with balanced gain and loss, show an intermediate decay rate , while Hermitian systems, lacking such protective mechanisms, decay fastest due to unmitigated decoherence. Physically, this reflects the interplay between temperature and decoherence at low temperatures. We observe a distinct effect on \( t_{\text{opt}} \), where the sub-Ohmic and super-Ohmic reservoirs exhibit an asymmetric behavior at low temperatures weaker thermal fluctuations (proportional to \( k_B T \)) result in slower decoherence, necessitating longer times to encode temperature information onto the qubit’s state via system-reservoir interactions \cite{A7}. At high temperatures, increased thermal noise accelerates decoherence, reducing \( t_{\text{opt}} \), and the environmental structure’s influence (captured by \( s \)) diminishes as thermal fluctuations dominate, leading to similar QFI behavior across all \( s \) values \cite{A8}. At low temperatures, however, the spectral density’s structure is critical, with sub-Ohmic (\( s = 0.5 \)) environments showing the most pronounced \( s \)-dependence due to strong low-frequency noise coupling. The lower plots in Fig.\ref{fig5}.(d-f) depict \( T_{\text{opt}} \), the temperature that maximizes the QFI for a given time \( t \) decreases as decoherence accumulates over time. APT sustains higher \( T_{\text{opt}} \) values longer, reflecting its ability to maintain sensitivity by reducing decoherence, which is crucial for small samples with limited interaction times, such as those using traveling qubits \cite{A4}. Physically, higher initial \( T_{\text{opt}} \) values indicate that the qubit can probe higher temperatures effectively at short times before decoherence erases quantum correlations, with the super-Ohmic regime (\( s = 3.0 \)) requiring higher temperatures due to weaker high-frequency coupling, while APT’s non-Hermitian dynamics consistently enhance performance across all regimes. This analysis demonstrates that optimizing QFI for temperature estimation hinges on the strategic manipulation of temporal and thermal parameters across varying symmetry conditions. Anti-PT (APT) symmetry emerges as the most effective, sustaining coherence and enhancing sensitivity through non-Hermitian dynamics, particularly in sub-Ohmic regimes with strong low-frequency coupling. These findings underscore the potential of APT symmetry to advance quantum thermometry, especially in constrained environments, offering a robust framework for future technological applications.

\section{Quantum thermometry of Ohmic-like samples}\label{sec5} 
\subsection{ Quantum information loss } 
\begin{figure*}[t] 
\includegraphics[width=5.5cm]{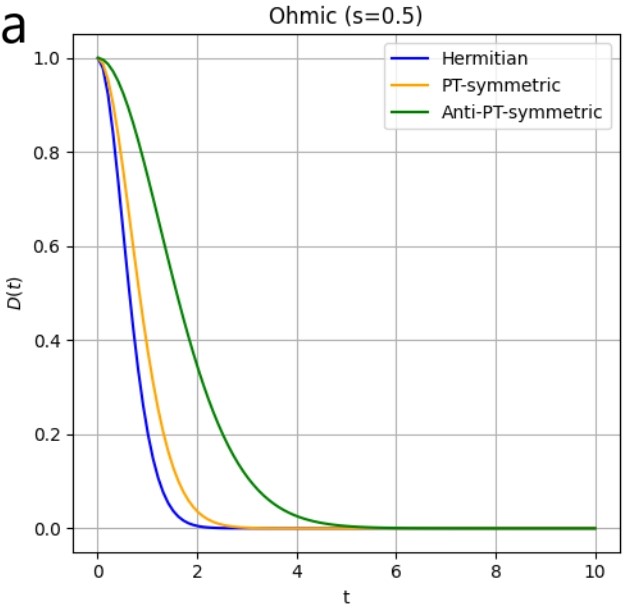}
     \includegraphics[width=5.45cm]{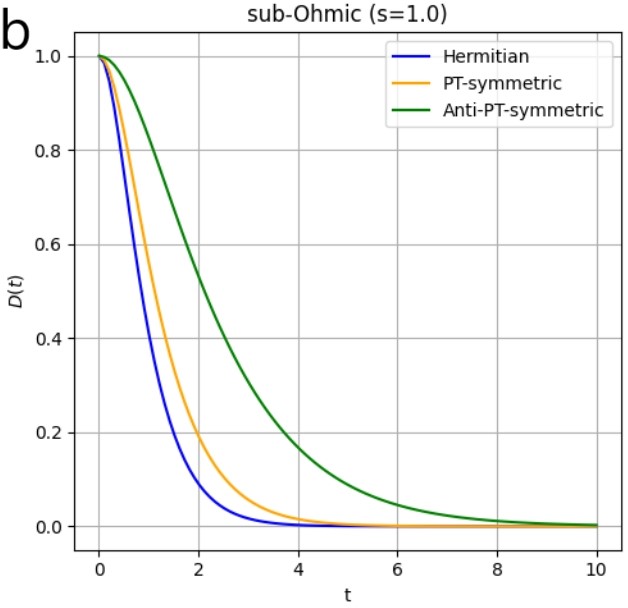}
\includegraphics[width=5.45cm]{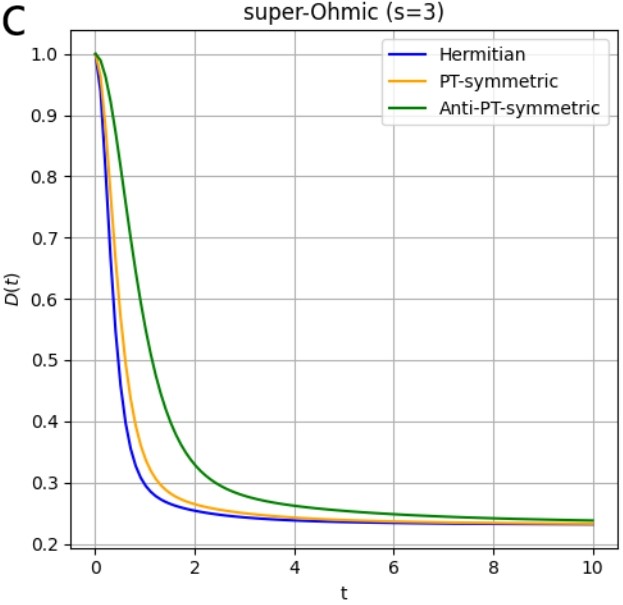}

      \caption{The decoherence  $D(T, t)$  as functions of the interaction time $t$ for three characteristic types of structured environments have been analyzed: (a) Ohmic reservoir ($s = 1$), (b) sub-Ohmic reservoir ($s = 0.5$), and (c) super-Ohmic reservoir ($s = 3$), under various symmetry conditions. The model \(D(t)=\exp(-t/a)\) was fitted by performing a linear regression of \(\ln D\) versus \(t\) over the region where \(D\) remains significant (the near-zero plateau, dominated by noise, was excluded); fitting the data for figure (a) yields \(a\approx 0.685\pm 0.019\) for the Hermitian (blue), \(a\approx 1.688\pm 0.066\) for the PT-symmetric (orange), and \(a\approx 3.141\pm 0.092\) for the Anti-PT-symmetric (green); the analysis indicates that the decoherence decay follows an exponential form \(D(t)\simeq e^{-t/a}\) in each case, and qualitatively similar fit results were obtained for figures (b) and (c); a larger constant \(a\) corresponds to a slower decay, so increasing \(a\) slows decoherence and better protects the system from information loss over time — concretely, the Hermitian trace (smallest \(a\)) decays fastest while the Anti-PT-symmetric trace (largest \(a\)) exhibits the slowest decay, indicating the greatest coherence protection.}
\label{fig6}
\end{figure*}

The analysis of decoherence and entropy dynamics reveals significant insights into the behavior of a quantum system coupled to an Ohmic reservoir (see Fig.\ref{fig6}.a-c). The symmetry effects demonstrate a pronounced influence, with the APT symmetry consistently exhibiting the slowest decoherence across all regimes ,  likely attributable to non-Hermitian dynamics near exceptional points that diminish environmental coupling. PT symmetry provides an intermediate level of protection through a balanced gain and loss mechanism, thereby reducing the effective decoherence rate \cite{A1}, whereas Hermitian dynamics, devoid of such protective mechanisms, display the fastest decoherence due to their complete exposure to environmental dissipation \cite{A3}. The noise spectrum plays a critical role, with the sub-Ohmic regime (\( s = 1.0 \)) inducing the most rapid decoherence owing to its dominant low-frequency noise, which aligns closely with the qubit's energy scale and enhances coupling . The Ohmic regime (\( s = 0.5 \)) presents a balanced yet still swift decoherence process, while the super-Ohmic regime (\( s = 3.0 \)) exhibits the slowest decoherence due to its weaker high-frequency coupling \cite{A8}. Regarding parameter contributions, the qubit's parameters (a, b, c, d) exert differential effects across symmetries. In the Hermitian case, a, b, and c augment decoherence, whereas in PT-symmetric systems, a and b increase it, though \(\theta\) acts to mitigate this effect. For the APT case, only a promotes decoherence, with b and c contributing to its reduction, thereby supporting its slower decay.

\begin{figure*}[t] 
\includegraphics[width=5.9cm]{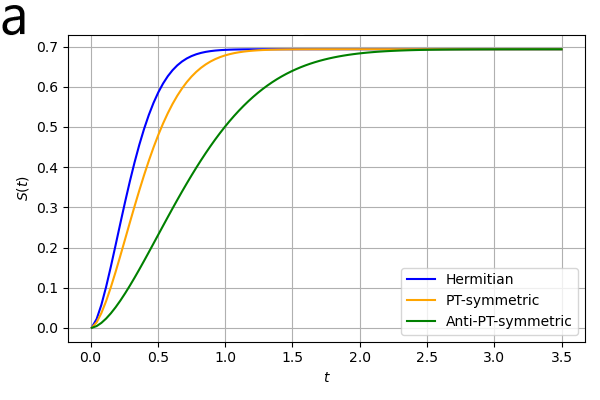}
        \includegraphics[width=5.9cm]{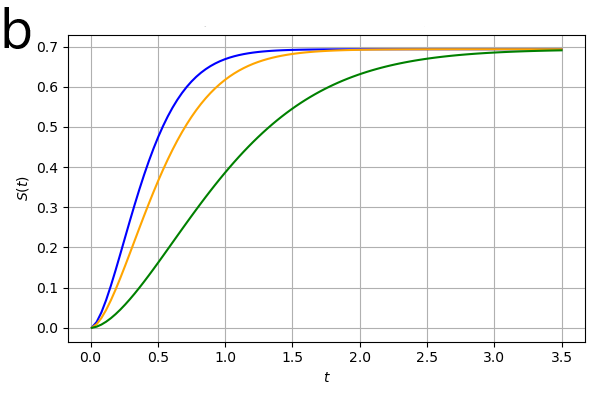}
        \includegraphics[width=5.9cm]{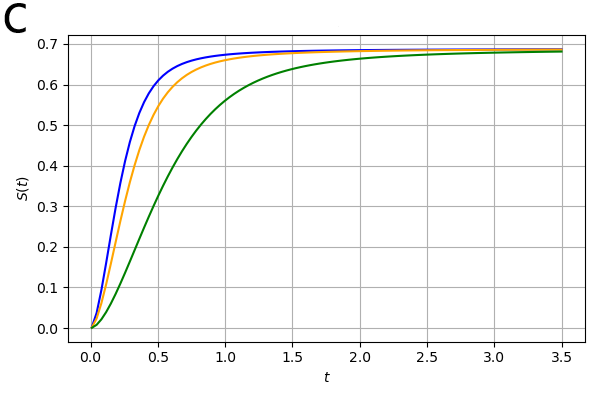}
      \caption{The von Neumann entropy $S(t)$ as functions of the interaction time $t$ for three characteristic types of structured environments have been analyzed: (a) Ohmic reservoir ($s = 1$), (b) sub-Ohmic reservoir ($s = 0.5$), and (c) super-Ohmic reservoir ($s = 3$), under various symmetry conditions.}
    \label{fig7}
\end{figure*}

To explore information loss, we analyze the correlation and information flow between the system and its environment through the von Neumann entropy \( S(t) \) over time in Fig.\ref{fig7}.(a-c). Initially, \( S(t) \) is zero reflecting a coherent qubit state, but rises rapidly due to decoherence from low-frequency noise (Ohmic spectrum \( s = 0.5 \)) as the system interacts with the reservoir. Hermitian dynamics show the steepest entropy increase, indicating strong dissipation without symmetry protection, compared to the slower rises in PT and APT symmetries. Intermediately, entropy growth persists as the qubit equilibrates, with Hermitian experiencing the most decoherence, PT showing a moderated rise due to balanced gain and loss, and APT exhibiting the slowest increase, likely due to non-Hermitian effects preserving coherence. At long times, \( S(t) \) plateaus around 0.65 across all symmetries, signifying thermal equilibrium, though APT's slightly lower entropy suggests reduced environmental entanglement, possibly from symmetry-stabilized eigenstates near exceptional points, reflecting maximum entanglement entropy influenced by the reservoir's temperature and spectral density. The observed trends underscore the pivotal role of symmetry in modulating decoherence within open quantum systems. Hermitian dynamics, lacking non-Hermitian protection, are highly susceptible to energy dissipation, resulting in rapid entropy growth. PT symmetry offers moderate resistance through gain-loss balance, whereas APT symmetry provides the strongest shield, potentially due to non-Hermitian phase transitions that mitigate environmental coupling. The Ohmic reservoir's dominance of low-frequency noise drives the system toward a thermal steady state, with the entropy plateau indicating an irreversible loss of quantum information into the environment, modulated by symmetry-dependent dynamics. These findings suggest that APT symmetry could be harnessed in quantum technologies to enhance coherence times in noisy environments. This investigation highlights the pivotal role of symmetry in modulating decoherence, with anti-PT symmetry offering the most robust protection against environmental dissipation through non-Hermitian dynamics. The observed entropy plateau and symmetry-dependent decoherence rates underscore the potential of APT symmetry to enhance coherence times, particularly in low-frequency-dominated Ohmic regimes. These findings advocate for its strategic application in advancing quantum technologies under challenging environmental conditions.
\begin{figure*}[t] 
        \includegraphics[width=5.9cm]{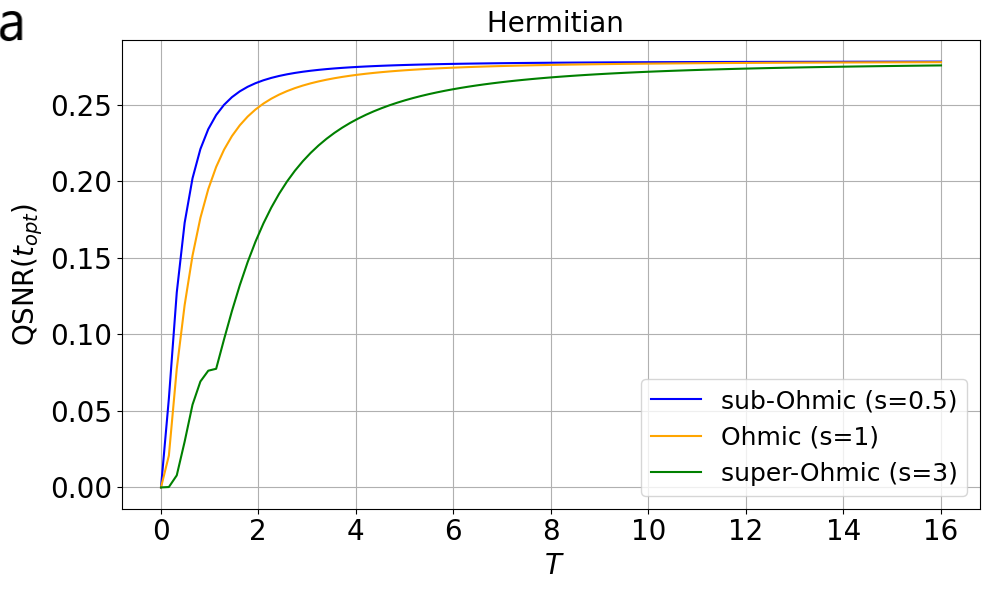}
        \includegraphics[width=5.9cm]{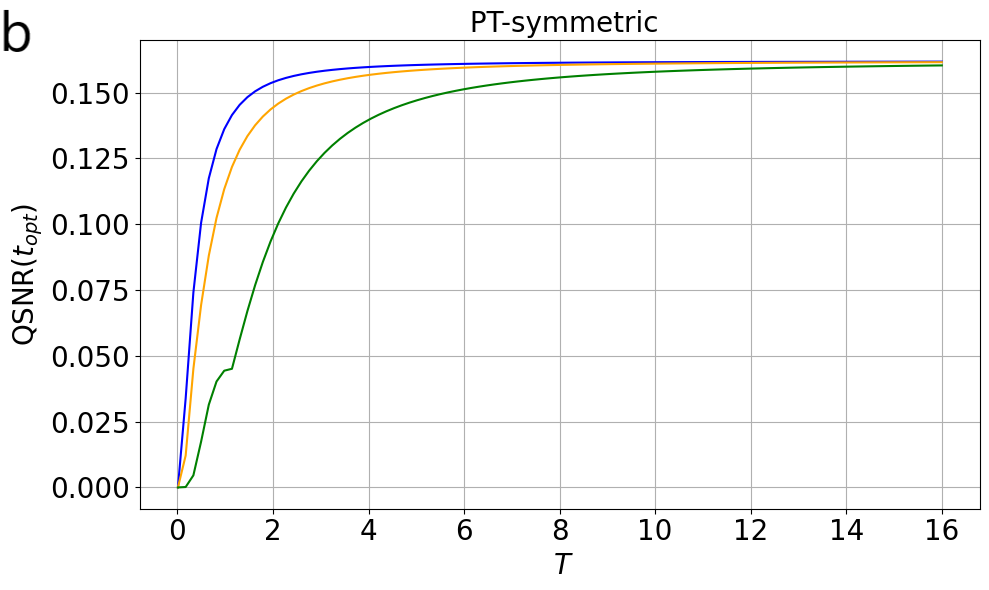}
        \includegraphics[width=5.9cm]{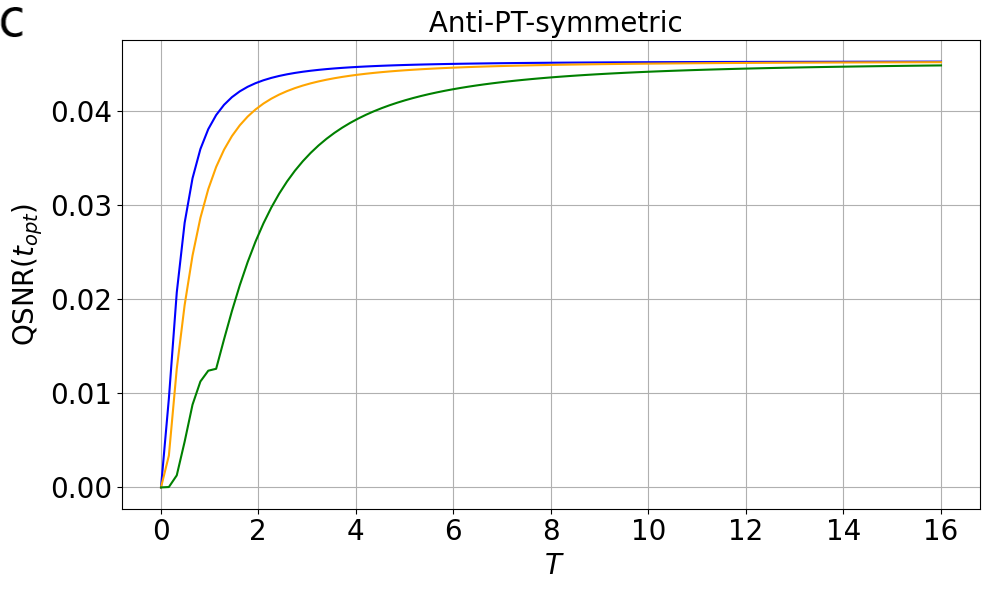}
      \caption{Plot of the QSNR evaluated at the optimal interaction time for different symmetry $t_{\text{opt}}$ as a function of the reservoir temperature $T$. Three classes of structured environments are considered: Ohmic ($s = 1$), sub-Ohmic ($s = 0.5$), and super-Ohmic ($s = 3$).}
    \label{fig8}
\end{figure*}
\subsection{Optimal Bound} 
The QSNR in Fig.\ref{fig8}.(a-c) is evaluated across the three distinct values of the Ohmicity parameter previously discussed. At low temperatures, the QSNR approaches zero, indicating a highly inefficient estimation process in this range. Conversely, as temperatures rise, the QSNR exhibits a noticeable increase. The response varies with \( s \) in the intermediate temperature range. At elevated temperatures, as noted earlier, the environmental structure becomes negligible, and the QSNR converges to a consistent universal value, irrespective of the environmental spectral density characteristics. For intermediate temperatures, it increases. The behavior is sdependent in the intermediate temperature regime, for higher $T$ the structure of the environment becomes irrelevant, and the QSNR saturates to a universal value, independent of the symmetry and the
nature of the spectral density of the environment. The influence of symmetry on the QSNR reveals that anti-PT-symmetric (APT) symmetry results in the lowest QSNR plateau at 0.04, likely attributable to non-Hermitian effects that diminish the effective coupling to the reservoir, thereby constraining temperature sensitivity. In contrast, PT symmetry yields a moderate QSNR plateau of 0.16, facilitated by a balance of gain and loss, while Hermitian symmetry achieves the highest QSNR at $0.25$, reflecting its complete exposure to thermal noise without any mitigating symmetry effects. Regarding the role of the noise spectrum, the sub-Ohmic regime (\( s = 0.5 \)) consistently generates the highest QSNR across all symmetries, driven by robust low-frequency noise coupling that enhances temperature estimation precision. Conversely, the super-Ohmic regime (\( s = 3 \)) exhibits the lowest QSNR due to the less effective coupling of high-frequency noise, which reduces sensitivity, while the Ohmic regime (\( s = 1 \)) strikes a balance, producing intermediate QSNR values. The temperature dynamics show that the initial rise in QSNR with \( T \) is a consequence of heightened thermal fluctuations, which strengthen the interaction between the qubit and the reservoir. This analysis reveals that the QSNR is significantly shaped by symmetry and noise spectrum, with Hermitian symmetry yielding the highest QSNR due to unmitigated thermal coupling, while APT symmetry offers the lowest due to reduced environmental interaction. The sub-Ohmic regime enhances precision through strong low-frequency coupling, contrasting with the less effective super-Ohmic regime. These findings underscore the potential for tailoring symmetry and spectral conditions to optimize temperature estimation in quantum systems.

\section{Conclusion}\label{sec6} 
In this study, we explored quantum thermometry using a single qubit subjected to dephasing, demonstrating its efficacy in accurately determining the temperature of Ohmic samples. Our approach is fundamentally quantum, leveraging the qubit's susceptibility to decoherence, with different symmetry classes of the system, including both Hermitian and non-Hermitian configurations. We observed that the QFI exhibits a peak as a function of interaction time at any given temperature for both Ohmic and sub-Ohmic samples, while for super-Ohmic samples, this peak manifests only at elevated temperatures. Consequently, optimal temperature estimation can be achieved at a finite interaction duration, prior to the qubit reaching its steady state. The exception occurs in super-Ohmic environments at low temperatures, where optimal estimation aligns with stationarity and thermalization, accompanied by a saturation phenomenon. Notably, preliminary insights suggest that non-Hermitian systems offer enhanced robustness against quantum decoherence, which could lead to improved thermometric performance under realistic noise conditions. Our findings indicate that achieving higher QFI values requires the probe to interact with the sample for a sufficient duration, thereby undergoing considerable coherence loss to effectively encode temperature information. However, optimal estimation does not always correspond to the regime of strong decoherence, as excessive decoherence may result in the loss of encoded information within the qubit state. Thus, the optimal conditions for temperature estimation arise from a complex interplay between the dephasing dynamics and the specific ohmic characteristics of the environment. Additionally, we noted that non-Markovian effects do not influence the temperature estimation process, as they are only significant in the dynamics of a probe interacting with a super-Ohmic environment at low temperatures, without contributing to improved estimation precision.

\section*{ACKNOWLEDGMENTS}

A.H acknowledges the financial support of the National Center for Scientific and Technical Research (CNRST) through the "PhD-Associate Scholarship-PASS" program. The authors acknowledge the LPHE-MS, FSR for the technical support.\par 

\textbf{Declaration of competing interest:}\par 
The authors declare that they have no known competing financial interests or personal relationships that could have appeared to influence the work reported in this paper.\par

\textbf{Data availability:}\par 
No data was used for the research described in the article.

\appendix

\end{document}